\documentclass{article}
\usepackage{hyperref}
\title{Kelvin's clouds
}
\author{Oliver Passon\footnote{University of Wuppertal, School for Mathematics and Natural Sciences, 42117 Wuppertal, Germany. passon@uni-wuppertal.de}
}
\date{\today}
\begin{document}

\maketitle

\begin{abstract}
 In 1900 Lord Kelvin identified two problems for 19th century physics, two``clouds" as he puts it: the relative motion of the ether with respect to massive objects, and Maxwell-Boltzmann's theorem on the equipartition of energy.  These issues were  eventually solved by the theory of special relativity and by quantum mechanics. In modern quotations, the content of Kelvin's lecture is almost always distorted and misrepresented.  For example, it is commonly claimed that Kelvin was concerned about the UV-catastrophe of the Rayleigh-Jeans law, while his lecture   actually makes  no reference at all to blackbody radiation.  I rectify these mistakes and explore reasons for their occurrence.   
\end{abstract}

\section{Introduction\label{intro}}
 In 1900, Lord Kelvin, born as William Thomson, delivered a lecture  in which he identified two problems for physics at the turn of the 20th century:  the question of the existence of ether and the equipartition theorem. The solution for these two clouds was eventually achieved by the theory of special relativity and by quantum mechanics \cite{kelvin1901}. Understandably, these prophetic remarks are often quoted in popular science books or in the introductory sections of textbooks on modern physics. A typical example is found in the textbook  by 
Tipler and Llewellyn \cite{tipler}:
\begin{quote}
 [...] as there were already vexing cracks in the foundation of what we now refer to as classical physics. Two of these were described by Lord Kelvin, in his famous Baltimore Lectures in 1900, as the ``two clouds" on the horizon of twentieth-century physics: the failure of theory to account for the radiation spectrum emitted by a blackbody and the inexplicable results of the Michelson-Morley experiment.
\end{quote}
Unfortunately, this and other citations of the lecture frequently misrepresent its content, as I will show in this paper.\footnote{Out of the thirteen quotations I could identify in the popular science and textbook literature using a Google scholar search, all were inaccurate in some way. For example, twelve of them erroneously mentioned the blackbody-radiation problem as one of Kelvin’s clouds. The Kelvin-story has now become part of the general folklore and is presumably often not referenced properly, or is quoted from the secondary literature (which perpetuates the common mistakes). 
By opposition, essentially all historically-oriented publications dealing with  Kelvin’s lecture portray it accurately.} Kelvin's lecture is briefly summarized in Sec.~\ref{whatsay}, before I deal with common misrepresentations and their origin in  Sec.~\ref{whatquote}. That textbooks contain inaccuracies and large or small mistakes is certainly no news.  However, I believe that this example  is particularly harmful for physics teaching. Additionally, studying it allows us to draw some general conclusions on the relation between the history of physics and physics teaching. This will be  briefly touched upon in  Sec.\ref{sum}.   

\section{Kelvin's lecture\label{whatsay}}
 On April 27, 1900, Lord Kelvin delivered the Friday Evening Lecture at the Royal Institution in London. An extended version of this lecture was submitted to the {\em Philosophical Magazine} in February 1901 and published in the July issue under the title ``Nineteenth Century Clouds over the Dynamical Theory of Heat and Light".  In this paper Lord Kelvin identified two urgent problems (``clouds") of ``the theory of heat and light". In his own words:
\begin{quote}
I. The first came into existence with the undulatory theory of light, and was dealt with by Fresnel and Dr. Thomas Young; it involved the question, How could the earth move through an elastic solid, such as essentially  is the luminiferous ether? II. The second is the Maxwell-Boltzmann doctrine regarding the partition of energy.
\end{quote}
Thus, the first cloud concerns the relative motion of the ether and ponderable bodies. Kelvin discussed various problems of mechanical ether models and also the (null)result of the Michelson-Morley experiment. While the suggestion of  FitzGerald and Lorentz to account for this finding by a contraction effect was mentioned, Kelvin left no doubt that he regarded ``cloud one"  as a grave problem. With typical British humor the paragraph concluding the ``cloud one'' discussion has only one sentence: ``\S 11. I am afraid we must still regard Cloud No. I. as very dense."

The following 45 (!) paragraphs are devoted  to the ``Maxwell-Boltzmann doctrine", i.e. the equipartition theorem, as we would say today.\footnote{According to the equipartition theorem, each degree of freedom contributes $\frac{kT}{2}$ to the mean energy if the system is in thermal equilibrium.} He questioned the derivation of this law (as he did before) and pointed also to its failure to explain certain specific heat  measurements. However, even where the equipartition theorem apparently predicts the specific heat correctly, Kelvin noted that it is at odds with the many spectral lines that indicate additional vibrational degrees of freedom. 

A long passage is devoted to -- as we would say today -- simulations of particle collisions in enclosed areas with irregular surfaces. Apparently, Kelvin's assistant William Anderson (whom Kelvin thanked in the introduction) generated random numbers by shuffling decks of numbered cards and calculated thousands of molecular impacts with surfaces and  inter-molecular collisions. Accordingly, Kelvin's paper is often cited in the computational physics literature and he gets credit for the first Monte Carlo simulation of a gas; see e.g. Ref. \cite{bird79}. Kelvin suggested that all these results support his previous arguments  and the concluding paragraph states that one should perhaps abandon the equipartition theorem.

\section{Kelvin's lecture in the literature\label{whatquote}}
 Kelvin's remark about the  two clouds is frequently cited, especially in popular science books and in  introductory sections of textbooks on modern physics. As early as 1938 the Russian-American physical chemist Saul Dushman (1883--1954) wrote in his textbook {\em The Elements of Quantum Mechanics} \cite{dushman}:
\begin{quote}
A little over a third of a century has passed since Lord Kelvin, in an address before the British Association, pointed out that there were apparently two clouds upon the scientific horizon. One of these was represented by the experiment of Michelson and Morley; the other involved the failure of classical theory in accounting for the observations on the energy distribution in the radiation emitted by a black body.
\end{quote}
Dushman confused here the  {\em Royal Institution} with the {\em British Association for the Advancement of Science}, which is a minor slip of pen. Note that the previous quotation from Tipler and Llewellyn (Sec.~\ref{intro}) refers to the ``famous Baltimore Lectures''. Apparently these authors got confused by the fact that the ``cloud-lecture'' was added to the reprint of the Baltimore Lectures in 1904, although the Baltimore lectures were delivered at Johns Hopkins University as early as 1884. While the location of the lecture is certainly of minor importance, these mistakes   indicate that the original source was not looked up. 

However, Dushman (like Tipler and Llewellyn above)   identified ``cloud two" with  the blackbody-radiation problem while Lord Kelvin did not mention this issue at all.\footnote{ However, Dushman's wording is ambiguous. He wrote that the other cloud ``involved'' the failure to describe the blackbody-radiation. Perhaps Dushman  was aware, that Kelvin's second cloud was the equipartition theorem. We should also make a brief comment on ``cloud one''. Its description as being related to the experiment of Michelson and Morley is fairly accurate. However, if textbook authors suggest (as they often do) that this finding prompted Einstein's theory of special relativity, this is incorrect. In fact,  this ether-drift experiment played no essential role for the genesis of Einstein's theory  at all (see Ref.~\cite{franklin}).}  This is perhaps the most common mistake encountered in the literature (see e.g. Ref.~\cite{alamri2016,deych2018,wilde2019} for other examples) and it is easy to explain.

Many textbooks portray Planck's discovery of his blackbody radiation law as the reaction to the so-called ``ultraviolet catastrophe'' of the Rayleigh-Jeans law.  It is true that the Rayleigh-Jeans law follows from the application of the equipartition theorem to the continuous radiation field (with its infinite   degrees of freedom) and diverges for high frequencies.  However, while Lord Rayleigh anticipated this law in a short note in 1900 \cite{rayleigh1900}, the Rayleigh-Jeans law was published only in 1905 \cite{jeans}. Planck's derivation of his celebrated radiation law in 1900 followed a completely different route and was unrelated to the ultraviolet divergence \cite{planck1900}. In brief, Planck reached his law in 1900 by modifying the derivation of Wien's radiation law (suggested in 1896; see Ref.~\cite{wien}) that he had published one year earlier (see Ref.~\cite{planck99}). The irrelevance of equipartition in Planck's work is discussed for example by Martin J. Klein and Helge Kragh \cite{klein,kragh}. In addition, Kelvin's lecture shows clearly that the equipartition theorem was no generally accepted part of ``classical physics'' which further contradicts the standard account of Planck's discovery.  In addition it is debated whether Planck introduced any energy quantisation at all (see e.g. Ref.~\cite{klein} and Ref.~\cite{passon}, which offer an overview of the debate on whether Planck's law implies quantum discontinuity.)     

Thus, to identify Kelvin's concern with respect to the equipartition theorem  with the blackbody-radiation problem is due to a conflation with this other common misconception regarding the origin of quantum theory. It is certainly true that quantum physics eventually resolved Kelvin's ``cloud two'' problem. This, however, was achieved not by Planck's work on blackbody radiation in 1900 but started with  Einstein's theory of specific heat in 1907 \cite{einstein1907} (which was certainly building on Planck's work). As noted for example by Helge Kragh (Ref.~\cite{kragh1999}, p. 69), it was this application of Planck's work that  played an important role in exciting the interest in Planck's energy quanta, since specific heat  was a more traditional field of physics. Hence, Kelvin was in fact pointing to the relevant problem for the later expansion of quantum physics and the story needs no distortion in order to be worth telling.     

In any event, the blackbody radiation problem was not even mentioned by Lord Kelvin in 1900. Even a distinguished scholar like David Bohm makes this blunder:
\begin{quote}  
For example, Lord Kelvin, one of the leading physicists of the time, expressed the opinion that the basic general outline of physical theories was pretty well settled, and that there remained only ``two small clouds” on the horizon, namely, the negative results of the Michelson-Morley experiment and the failure of Rayleigh-Jeans law to predict the distribution of radiant energy in a black body. It must be admitted that Lord Kelvin knew how to choose his ``clouds”, since these were precisely the two problems that eventually led to the revolutionary changes in the conceptual structure of physics that occurred in the twentieth century in connection with the theory of relativity and the quantum theory. \cite{bohm}
\end{quote}
Here, however, one encounters another typical misrepresentation of Lord Kelvin's lecture, namely that he allegedly just mentioned two ``small'' clouds, thus downplaying the importance Lord Kelvin seems to give to these issue. A similar sentiment is expressed by Willem de Muynck. He was not committing the ``blackbody-fallacy'' but wrote:
\begin{quote}
Lord Kelvin saw in 1901 only two small ``clouds" in the sky, namely, the ``anomalous" behavior of specific heat at low temperature and the Michelson-Morley experiment, but he was convinced that an explanation could be found for every difficulty. \cite{demuynck2002}
\end{quote}
In fact, nowhere in his presentation did Kelvin belittle the threat or call the clouds ``small'' (he did not call them ``dark" either -- as for example Rechenberg claims in Ref.~\cite{rechenberg2010}). Kelvin also did  not suggest that these problems could be solved quickly. With respect to the ether problem I have already quoted,  his assessment was  that this cloud seemed ``very dense''. With respect to the equipartition theorem, he likewise expressed an overall puzzlement.

The origin of this misconception seems again related to another misquote, namely Kelvin's alleged claim that physics had essentially reached its end. Note that Bohm in the above quotation  expressed the same sentiment. Another example can be found in a recent book by  the prominent science writer Brian Clegg \cite{clegg2014}:
\begin{quote}
The hubris  of  the  scientific establishment is probably best summed up by the words of  a  leading physicist of  the  time, William Thomson, Lord  Kelvin. In 1900 he commented, no doubt in rounded, selfsatisfied tones: ``There is nothing new to be discovered in physics. All that remains is more and more precise measurement." (p. 15)
\end{quote}
This statement is also ``quoted" by the eminent physicist  Muhammad S. Zubairy \cite{alamri2016}:
\begin{quote}
According to a quote, attributed to Lord Kelvin in an address to the British Association for the Advancement of Science in 1900, ``There is nothing new to be discovered in physics now. All that remains is more and more precise measurement." [...] There were, however, two “clouds” on the horizon of physics at the dawn of the twentieth century. 
\end{quote}
However, the claim that ``there is nothing new to be discovered in physics" is nowhere to be found in the 1900 lecture, or in any of Kelvin's other writings, and Zubairy suspiciously called this quotation   just ``attributed'' to Lord Kelvin. In fact, it has been suggested that it is  a similar quotation by Albert A. Michelson (namely ``The more important fundamental laws and facts of physical science have all been discovered [...]'' in Ref.~\cite{aam}, p. 23) that is misattributed to Kelvin.\footnote{See \url{https://en.wikiquote.org/wiki/William_Thomson}. However, this page tries also to correct the claim that the 1900 lecture was delivered at the {\em British Association} and locates it at the {\em Royal Society} instead of the {\em Royal Institution}. This problematic quote is also discussed by Javier Yanes  in this blog post: \url{https://www.bbvaopenmind.com/en/science/physics/lord-kelvin-and-the-end-of-physics-which-he-never-predicted/}.}

All this shows, that one is dealing with a narrative that has been carefully crafted in order to emphasize the rift between ``classical'' and ``modern'' physics. Against the backdrop of something considered essentially complete (with just two small blemishes), the conceptual changes of modern physics appear even more sensational and dramatic. This somewhat rhetorical  function is further strengthened  by calling Lord Kelvin  a ``leading physicist of the time'' (Bohm, Clegg). In a similar vein, Menas Kafatos and Robert Nadeau wrote \cite{kafatos90}:
\begin{quote}
Toward the end of the nineteenth century, Lord Kelvin, one of the best known and most respected physicists at that time, commented that ``only two small clouds'' remained on the horizon of knowledge in physics. 
\end{quote}
It is certainly true that Lord Kelvin was a highly respected and well known figure.  Especially in Britain he was famous for his work on the Atlantic telegraph (1866), for which he received his knighthood in 1892, and the very fact that he was invited to deliver the prestigious Friday Evening Lecture at the Royal Institution shows his high reputation.  However, by the time of the lecture in 1900 (at the age of 75), his lasting  contributions to thermodynamics and electromagnetism had been made decades ago. By that time,  Kelvin was widely regarded  in the community   as the last prominent adherent of a mechanical world view. Ole Knudsen in Ref.~\cite{knudson} remarked that Kelvin's ``refusal to accept Maxwell's concept of displacement current and his insistence that electrostatic action had to be propagated as condensational waves in the ether  had left him hopelessly behind the newer trend in physics.''\cite{bl1} These ``newer trends'' are characterized as the renunciation of   dynamical models in favor of  mathematical structures (See also Wise and Smith in Ref.~\cite{b87}). When Kelvin published the revised version of his Baltimore Lectures in 1904 (including the ``cloud-lecture'' in Appendix B), the reaction seemed  (to quote Knudson again) ``polite indifference, and it is hard to point to a physicist whose work was influenced by the book'' \cite{knudson}.

As noted by Knudson, Kelvin was opposed to the so-called electromagnetic world view that had become increasingly fashionable around the turn of the century \cite{darrigol}. This  view tried to base all of physics on electromagnetism. It is thus wrong to portray the advent of quantum physics as the overturn of the purely mechanical  classical physics, since this mechanical view was under attack already.  As pointed out by Richard Staley, the label of ``classical physics'' overemphasizes the uniformity of 19th century physics \cite{staley}.  In fact, Staley argued that the very notion of ``classical physics'' was co-created together with the notion of ``modern physics''  in 1911 and also served   a strategic function. 

\section{Concluding remarks\label{sum}}
 In 1960, the Philosophical Society of Washington asked Edward Condon to speak about   ``60 Years of Quantum Physics'' in his retiring presidential address. This was a fitting choice, given that Condon was a pioneer of quantum physics himself. After finishing his PhD at Berkeley, he had spent the fall of 1926 in G\"ottingen working with Max Born, and the spring of 1927 in Munich working with Arnold Sommerfeld. Thus, he was even an eyewitness  to some  of these exciting developments to which he also contributed. 

His lecture on the 1500th Regular Meeting of the Society took place at the Natural History Museum Auditorium of the Smithsonian Institution in Washington, D.~C. on December 2, 1960. Condon  started with Planck's discovery of  quantization in 1900 (published on December 14, 1900, hence the title ``60 Years...''). When he turned to Einstein's light quantum hypothesis, he confessed with  disarming honesty \cite{condon}: 
\begin{quote}
Let us now turn to Einstein's famous 1905 paper, which I must confess I had not read until I got to thinking over the preparation for this lecture. It is one of the papers we all hear about in school and worship, but do not read.
\end{quote}
He continues to express his astonishment that Einstein's paper does not contain Planck's constant $h$ (``believe it or not''). Apparently, Condon was not aware that Einstein's derivation was solely based on Wien's radiation law probably due to the fact that most textbooks claim that he applied Planck's law. Ironically, Condon was a textbook writer himself and, together with Philip Morse,  had authored one of the first textbooks on quantum mechanics in 1929. In this book, one finds with respect to Einstein's light quantum and  photoelectric effect ($W$ denotes the energy of the light): ``Einstein, in 1905, applied Planck's assumptions to the phenomenon and showed that $W$ must equal $h\nu$'' (Ref.~\cite{condon29}, p. 3).  

This little anecdote illustrates some of the reasons why science textbooks are famous for their poor accounts of history (see e.g. Ref. \cite{franklin,klein72,kragh92}).  Whitaker once coined the term ``quasi-history" for these simplifying and distorting narratives found in textbooks \cite{whitaker79}, and more recently Douglas Allchin has criticized what he calls ``myth-conceptions" in the textbook literature \cite{allchin}. 


The eminent historian of physics Allan Franklin believes that it is not necessarily a bad thing if  the history presented in physics textbooks is often inaccurate. He argued that,  after all, the purpose of these books is  to help students learn physics and ``[a]n inaccurate history may serve a pedagogical purpose''. \cite{franklin} However, Franklin believes that those who teach should know the ``accurate history'' and argues that also students should learn (the ultimately historical lesson) that science is not an ``unbroken string of successes''. At stake is not just historical accuracy but an adequate picture of scientific practice:
\begin{quote}
It is important, however, to present students with an accurate picture of the practice of science. This should include the fallibility of science. Too many textbooks  show  only  the  successes  of  science,  with  little  or  no  mention  of  the failures \cite{franklin}.
\end{quote}
This is the well known complaint that the quasi-historical narratives in textbooks  present science as a cumulative sequence of events which ultimately leads to the acceptance of the current theories while ignoring failures, social and political context or the original motivation of the respective scientist \cite{whitaker79,brush74}.


What makes the story of Kelvin's clouds special is that  the ``failure" of physics {\em is} highlighted and that those who misquote him are not trying to present the advent of  quantum physics or relativity as part of an ``unbroken string of successes''. On the contrary,  we are presented with the  ``leading physicist'' Lord Kelvin who allegedly  believed in the almost finality of ``classical physics'', while pointing unconsciously to precisely {\em the} two problems which led to the   upheaval of physics within a short time.  To emphasize the revolutionary nature of the theories of relativity and quantum mechanics, and to downplay the ways that they are more continuously linked to the theories that preceded, might also be a reaction to a widespread (but simplistic) understanding of Kuhn's notions of scientific revolution and paradigm shift.\footnote{ I thank one of the referees for suggesting this connection to me. For an overview about the impact of Kuhn's work on physics education see Ref.~\cite{kuhn}.}

That is, when textbook history addresses failure it falls into the other extreme and portrays it as {\em complete failure}. Apparently (and unfortunately), quasi-history knows only {\em complete success} or {\em complete failure} and nothing in between.


\end{document}